\pdfoutput=1

\documentclass[aps,twocolumn,amsmath,amssymb,preprintnumbers,superscriptaddress,floatfix]{revtex4}
\setcitestyle{numbers,square,sort,compress}
\usepackage[utf8]{inputenc}
\usepackage{newtxtext}
\usepackage[upint]{newtxmath}
\usepackage{microtype}
\usepackage{textcomp}
\usepackage{dsfont}
\usepackage{eucal}
\usepackage{bm} 

\usepackage{enumerate}
\usepackage{amsfonts}
\usepackage{amsmath}
\usepackage{amssymb}
\usepackage{braket}
\usepackage{color}
\usepackage{soul}

\usepackage{graphicx}

\usepackage[colorlinks,allcolors=blue]{hyperref}
\usepackage[capitalize]{cleveref}

\newcommand{\prlsection}[1]{\textit{#1}.---\hspace{0.05em}}

\let\phi=\varphi

\renewcommand{\d}[1]{\mathrm{d}{#1}}                            % Differential scalar element
\definecolor{DarkRed}{rgb}{0.80,0,0}
\definecolor{Purple}{rgb}{0.55,0,0.55}

\DeclareMathOperator{\im}{Im}

\renewcommand{\d}[1]{\mathop{\textrm{d}\kern0em#1\kern0.1em}}
\usepackage{relsize}
\newcommand{\gr}{\H{g}^{\textrm{\smaller R}}{}} 
\newcommand{\ga}{\H{g}^{\textrm{\smaller A}}{}}
\newcommand{\gk}{\H{g}^{\textrm{\smaller K}}{}}
\newcommand{\gh}{\H{h}}

\newcommand{\ir}{\H{I}^{\kern0.07em\textrm{\smaller R}}}
\newcommand{\ia}{\H{I}^{\kern0.07em\textrm{\smaller A}}}
\newcommand{\ik}{\H{I}^{\kern0.07em\textrm{\smaller K}}}

\renewcommand{\H}[1]{\hat{#1}}
\newcommand{\V}[1]{\check{#1}}
\newcommand{\B}[1]{\bm{#1}}

\newcommand{\BH}[1]{\H{\B{#1}}}

\newcommand{\up}{\uparrow}                                      % Spin up
\newcommand{\dn}{\downarrow}                                    % Spin down
\begin{document}
\title{Controlling spin supercurrents via nonequilibrium spin injection}
\author{Jabir Ali Ouassou}
\affiliation{Center for Quantum Spintronics, Department of Physics, Norwegian \\ University of Science and Technology, NO-7491 Trondheim, Norway}

\author{Jason W. A. Robinson}
\affiliation{Department of Materials Science and Metallurgy, University of Cambridge, \\ 27 Charles Babbage Road, Cambridge CB3 0FS, {United Kingdom}}

\author{Jacob Linder}
\affiliation{Center for Quantum Spintronics, Department of Physics, Norwegian \\ University of Science and Technology, NO-7491 Trondheim, Norway}
 
\begin{abstract}
  We propose a mechanism whereby spin supercurrents can be manipulated in superconductor/ferromagnet proximity systems via nonequilibrium spin injection.
  {We} find that if a {spin supercurrent} exists in equilibrium, a nonequilibrium {spin accumulation} will exert a torque on the spins transported by this current.
  This interaction causes a new {spin supercurrent contribution} to manifest out of equilibrium, which is {proportional to and} polarized perpendicularly to both the injected spins and equilibrium spin current.
  This is interesting for several reasons: as a fundamental physical effect; due to possible applications as a way to control spin supercurrents; and timeliness in light of recent experiments on spin injection in proximitized superconductors.
\end{abstract}

\maketitle

\prlsection{Introduction}%
In the {field} of superconducting spintronics, {a key} objective is to study the interactions between superconductors~(S) and ferromagnets~(F) \cite{Linder2015,Eschrig2010,Eschrig2015a,Blamire2014}.
These interactions {produce} new types of Cooper pairs $\ket{\up\up}$~and~$\ket{\dn\dn}$ with {a} net spin polarization, which enables the use of S/F systems for dissipationless spin transport. 
{From a fundamental physics perspective, such an interplay between different types of quantum order is expected to give rise to interesting physics to explore.
Ultimately, the ambition is to exploit the interactions in S/F systems in devices related to {e.g.\ supercomputing} and ultrasensitive detection of heat and radiation.}

{One particular area that has attracted attention is how one may generate controllable} spin supercurrents in equilibrium, either via inhomogeneous magnetism~\cite{Grein2009,Alidoust2010,Shomali2011, Brydon2011, Moor2015,Gomperud2015a} or spin--orbit coupling~\cite{Konschelle2015, Jacobsen2016, Montiel2018, Bobkova2004}.
In the magnetic case, it has been shown that {two layers with noncollinear magnetic moments} $\B{m}_{{1}}$ and $\B{m}_{{2}}$ give rise to an equilibrium spin supercurrent $\B{J}_\text{s}^\text{eq} \sim \B{m}_{{1}} \times \B{m}_{{2}}$~\cite{Slonczewski1989}.
{While most work so far relies on magnetic control of spin supercurrents via rotation of $\B{m}_{{1}}$ relative {to} $\B{m}_{{2}}$, it would be interesting
%One challenge with these suggestions is{,} however{,} that they require magnetic
%would be difficult to manipulate the polarization of the spin supercurrent \textit{in situ}.
%To produce all spin projections, one would need to carefully control the magnetic orientations $\B{m}_{{1}}$ and $\B{m}_{{2}}$ independently.
%{In addition,} this would likely lead to high switching times for prospective device applications.
%It would therefore be interesting 
to determine if a spin supercurrent can be controlled via electronic spin injection instead.}
%\st{This would have the added benefit of enabling the coupling of} 
{Such a mechanism might be more beneficial with regard to coupling} {superconducting and nonsuperconducting} spintronics devices.
Note that this is different from many previous works on spin injection in superconductors, which were largely explained in terms of quasiparticles and not a spin-triplet condensate~\cite{Beckmann2016, Quay2013, Wakamura2014, Hubler2012, Silaev2015, Bobkova2015a, Bobkova2016}.

Recently, there has been a renewed interest in using nonequilibrium spin injection as a {means} to manipulate spin supercurrents.
This is largely due to a recent spin-pumping experiment~\cite{Jeon2018}, where microwaves were used to excite spins in the ferromagnetic layer of a{n} N/S/F/S/N junction.
{{This experiment showed that} the pumped spin current could increase below the critical temperature~$T_\text{c}$ of the S~layers. It has been proposed that this effect was assisted by a Cooper-pair spin supercurrent \cite{Montiel2018}, although alternative explanations have been proposed~\cite{TairaArxiv}.} 
%Depending on the spin--orbit coupling in the normal metal~N, {the spin current leaving F} could increase below 
%The main interpretation proposed in that paper was that the increased spin transmission for temperatures~$T<T_\text{c}$ was due to a spin-triplet supercurrent, although alternative explanations have been proposed~\cite{TairaArxiv}.
%We will not consider {that} specific geometry {here}, but}
%{Here, we} {focus on} the more general problem of how spin injection \st{might} affect spin supercurrents.
%One approach was explored in Ref.~\cite{Montiel2018}, where they considered Fermi-liquid interactions.
%In their model, a spin accumulation~$\B{\rho}_{\text{s}}$ induces an effective magnetic exchange field~$\B{m}_{\text{eff}} \sim \B{\rho}_\text{s}$.
%{Since} inhomogeneous exchange fields are well-known to produce spin supercurrents, this provides one mechanism for the generation of spin supercurrents.

\begin{figure}[t!]
  \includegraphics[width=\columnwidth]{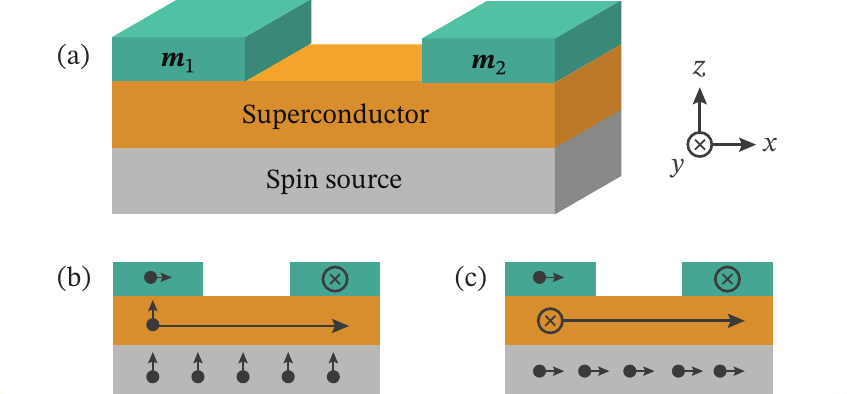}
  \caption%
  {%
    (a)~Magnetic insulators with magnetizations $\B{m}_{{{1}}}$ and $\B{m}_{{{2}}}$ {on a superconductor}.
    In equilibrium, this yields a spin supercurrent~$\B{J}_\text{s}^\text{eq} \sim \B{m}_{{{1}}} \times \B{m}_{{{2}}}$.
    A spin source injects a spin accumulation~$\B{\rho}_\text{s}$, which exerts a torque on the spins transported by the equilibrium current, resulting in a new contribution~${\B{J}_{\text{s}}^\text{neq} \sim \B{\rho}_\text{s} \times (\B{m}_{{{1}}} \times \B{m}_{{{2}}})}$.
    (b)~If the magnets are {magnetized} in the $x$- and $y$-directions, an equilibrium spin-$z$ supercurrent {arises}.
    Injection of spin-$z$ particles does not affect its polarization.
    Note that a spin supercurrent is in general a rank-2 tensor, which encodes both a polarization direction (short arrow) and transport direction (long arrow).
    (c)~If spin-$x$ particles are injected, however, a new spin-$y$ supercurrent component is generated.
    Similarly, spin-$y$ injection would {produce a spin-$x$ component.}
  }
  \label{fig:model}
\end{figure}

In this manuscript, we {consider how an injected nonequilibrium spin accumulation in general affects a{n existing} spin supercurrent (see \cref{fig:model})}.
We show that {such a spin injection} actually produces new terms in the equations for the spin supercurrent itself.
These terms have a natural interpretation in the form of the injected spin accumulation~$\B{\rho}_\text{s}$ exerting a torque on an equilibrium spin supercurrent~$\B{J}_\text{s}^\text{eq}$, thus giving rise to a new component $\B{J}_\text{s}^\text{neq} \sim \B{\rho}_\text{s} \times \B{J}_\text{s}^\text{eq}$ perpendicular to both. 
{Although this term occurs out-of-equilibrium, it shares the property of an equilibrium spin supercurrent that it does not require {gradients in any} chemical potential.
Therefore, it is legitimate to refer to the new term~$\boldsymbol{J}_\text{s}^\text{neq}$ as a supercurrent flowing without dissipation, as there is no energy loss  associated with a spatially varying chemical potential.}
{Our result is different from e.g.\ Ref.~\cite{Montiel2018}, which proposed that \emph{equilibrium} spin accumulation might produce spin supercurrents in some materials.}
%We note that our mechanism differs from the one in Ref.~\cite{Montiel2018} in three important ways.
%{Our} mechanism cannot cause a spin supercurrent to appear in a system without an equilibrium spin supercurrent since $\B{J}_\text{s}^\text{eq} \rightarrow 0 \Rightarrow \B{J}_\text{s}^\text{neq} \rightarrow 0$.
%{It is also} more universal as it does not require materials with specific Fermi-liquid parameters (\textit{i.e.}, metals near the Stoner criterion){, and the effect} is relevant for spin injection in any superconducting system with equilibrium spin transport.
%Finally, the spin supercurrent in Ref.~\cite{Montiel2018} is manipulated via an \emph{equilibrium} spin accumulation, while our result is inherently {nonequilibrium}.

\prlsection{Analytical results}%
Let us first consider a material with a spin-independent density of states~$N(\epsilon)$, where $\epsilon$ is the quasiparticle energy.
The nonequilibrium spin accumulation~$\B{\rho}_{\text{s}}$ can then be related to a spin distribution function~$\B{h}_\text{s}$,
\begin{equation}
  \B{\rho}_\text{s} = -\frac{\hbar}{2} \int_0^\infty\d\epsilon\, N(\epsilon)\,\B{h}_\text{s}(\epsilon),
\end{equation}
where $\B{h}_\text{s}$ describes the imbalance between spin-up and spin-down occupation numbers.
We define $\B{h}_\text{s}$ as a vector that points in the polarization direction of the spins, while its magnitude can be described in terms of a \emph{spin voltage}~$V_\text{s}$,
\begin{equation}
  h_\text{s}(\epsilon) = \big\{ \tanh[(\epsilon+eV_\text{s})/2T] - \tanh[(\epsilon-eV_\text{s})/2T] \,\big\}/2,
\end{equation}
where $e<0$ is the electron charge and $T$ is the temperature.
The spin voltage is defined as~$V_\text{s} \coloneq (V_\up - V_\dn)/2$, where $V_\sigma$ are the effective potentials seen by spin-$\sigma$ quasiparticles~\cite{OuassouArxivA,BergeretArxiv,Bauer2012}, {and $\boldsymbol{h}_\text{s}$ defines the spin-quantization axis}.

For a normal metal at $T=0$, the density of states ${N(\epsilon) = N_0}$ is flat, while the spin distribution $|\B{h}_{\text{s}}| = 1$ for ${|\epsilon| < eV_\text{s}}$.
This results in a spin accumulation $|\B{\rho}_\text{s}| = (\hbar/2) N_0eV_\text{s}$ that increases linearly with {$V_\text{s}$}.
This gives a simple interpretation of {$V_\text{s}$} as a control parameter: if the spin source in \cref{fig:model} is a nonsuperconducting reservoir, the spin voltage~{$V_\text{s}$} is directly proportional to the spin accumulation {in the reservoir}.

Similarly to the above, the excitation of quasiparticles from the Fermi level is decribed by an energy distribution~$h_0(\epsilon)$,
\begin{equation}
  \label{eq:energy-dist}
  h_\text{0}(\epsilon) = \big\{ \tanh[(\epsilon+eV_\text{s})/2T] + \tanh[(\epsilon-eV_\text{s})/2T] \,\big\}/2.
\end{equation}
At low temperatures, this shows that a spin voltage~$V_\text{s}$ also excites quasiparticles in a region of width~$2eV_\text{s}$ around the Fermi level~$\epsilon = 0$.
For a more in-depth discussion of the nonequilibrium distribution function, see Refs.~\cite{OuassouArxivA,BergeretArxiv,Silaev2015}.

Spin supercurrents can in general be expressed as {energy integrals of spectral spin supercurrents},
\begin{equation}
  \B{J}_\text{s} = -\frac{\hbar}{2} N_0 \int_0^\infty\d\epsilon\,{\im\!\left[\B{j}_\text{s}\right]}.
  \label{eq:spincurrent-integral}
\end{equation}
In equilibrium, the spectral current $\B{j}_\text{s}^\text{eq}$ is given by~\cite{Gomperud2015a,Eschrig2015a}
\begin{equation}
  \B{j}_\text{s}^\text{eq} = {\big( \B{g}_t \times \nabla \B{g}_t - \B{f}_t \times \nabla \tilde{\B{f}}_t \big) h_0}.
  \label{eq:j-eq}
\end{equation}
Here, $\B{g}_t$ describes the spin-polarization of the density of states, while $\B{f}_t$ describes spin-triplet correlations~\cite{Eschrig2015a}.
The cross products should be taken between the orientations of the vectors $\B{g}_t$ and $\B{f}_t$.
Since the gradient of a vector is a rank-2 tensor, the spin supercurrent {is} such a tensor, enabling it to encode both the spin polarization and the transport direction.
However, in effectively 1D systems like \cref{fig:model}, we can let the position derivative $\nabla \rightarrow \partial_x$.
In other words, in systems with 1D transport, the spin supercurrent reduces to a vector that describes spin polarization.
Note that the result depends only on the energy distribution~$h_0$, which is the only part of the distribution function which remains finite in equilibrium.

Outside equilibrium, the spin distribution~$\B{h}_\text{s}$ can become finite, and the spectral current gains an additional contribution:
\begin{equation}
  \B{j}_\text{s}^\text{neq} = {\big( \B{g}_t \times \nabla \B{g}_t - \B{f}_t \times \nabla \tilde{\B{f}}_t \big) \times i\B{h}_\text{s}}.
  \label{eq:j-neq}
\end{equation}
The full derivation of this result is included in the Supplemental information.
{The structure of \cref{eq:j-neq} is very reminiscent of \cref{eq:j-eq}, since both depend on $\B{g}_t \times \nabla \B{g}_t - \B{f}_t \times \nabla \tilde{\B{f}}_t$.}
However, {its} cross product structure {generates} a spin current \emph{perpendicular} to the one in \cref{eq:j-eq}.
We also see that it {contains an extra factor~$i$; since the distribution functions $h_0$ and $\B{h}_\text{s}$ are both real, this causes \cref{eq:spincurrent-integral} to extract the real and not imaginary part of $\B{g}_t \times \nabla \B{g}_t - \B{f}_t \times \nabla \tilde{\B{f}}_t$.
This comparison shows that the nonequilibrium contribution can be summarized as}
\begin{equation}
  \B{j}_\text{s}^\text{neq} = \B{j}_\text{s}^\text{eq} \times ({i}\B{h}_\text{s}/h_0).
  \label{eq:j-neq-eq}
\end{equation}
So long as $\boldsymbol{g}_t \times \nabla \boldsymbol{g}_t - \boldsymbol{f}_t \times \nabla \boldsymbol{f}_t$ is a complex number---which it in general is---it produces an equilibrium spin supercurrent $\boldsymbol{j}_\text{s}^\text{eq}$ according to \cref{eq:j-eq,eq:spincurrent-integral}, which combined with a finite {spin distribution} $\boldsymbol{h}_\text{s}$ immediately produces the new supercurrent term $\boldsymbol{j}_\text{s}^\text{neq}$ according {to \cref{eq:j-neq-eq}.}
This suggests an intuitive interpretation of the effect: injected spins described by $\B{h}_\text{s}$ exert a torque on the spins transported by the equilibrium component~$\B{j}_\text{s}^\text{eq}$, producing a nonequilibrium component $\B{j}_\text{s}^\text{neq}$ perpendicular to both.
This result also proposes that this nonequilibrium spin supercurrent should increase linearly with the equilibrium spin supercurrent and the injected spin accumulation.
{Thus, an equilibrium spin supercurrent gains a new component when propagating through a region with spin accumulation $\boldsymbol{\rho}_\text{s}$.}
{All these predictions that arise from \cref{eq:j-neq-eq} are confirmed numerically later in this paper.}

Let us now consider the setup {in} \cref{fig:model}.
In equilibrium, the $x$- and $y$-polarized magnets give rise to a $z$-polarized spin supercurrent $\B{j}_\text{s}^\text{eq} \sim \B{z}$.
A generic spin source then introduces a spin imbalance in the superconductor, which we describe via a nonzero spin distribution~$\B{h}_\text{s}$.
If these spins are polarized in the $z$-direction, meaning that $\B{h}_\text{s} \parallel \B{j}_\text{s}^\text{eq}$, then the nonequilibrium contribution $\B{j}_\text{s}^\text{neq} = 0$.
On the other hand, if these spins are polarized in the $x$-direction, so that $\B{h}_\text{s} \perp \B{j}_\text{s}^\text{eq}$, then the nonequilibrium contribution $\B{j}_\text{s}^\text{neq} \sim \B{j}_\text{s}^\text{eq} \times \B{h}_\text{s} $ obtains a $y$-polarized component proportional to the spin imbalance.
Similarly, if one had injected spin-$y$ particles instead, a spin-$x$ supercurrent would appear in the superconductor.

To summarize, for the geometry in \cref{fig:model}, the analytical results suggest that we should expect a spin-$y$ supercurrent proportional to the spin-$x$ voltage, while the spin supercurrent should remain unchanged for a spin-$z$ voltage.
In the following sections, we compare these expectations to numerical results.

\prlsection{Technical details}%
We perform the numerical calculations using the Usadel equation~\cite{BergeretArxiv,Chandrasekhar2008,Belzig1999,Rammer1986,Usadel1970}, which describes superconducting systems in the quasiclassical and diffusive limits.
Within this formalism, observables are described via $8\times8$ quasiclassical propagators in Keldysh\,$\otimes$\,Nambu\,$\otimes$\,spin space,
\begin{equation}
  \V{g} \coloneq
  \begin{pmatrix}
    \gr & \gk \\
    0   & \ga
  \end{pmatrix}.
\end{equation}
These components are related by the identities $\gk = \gr \gh - \gh \ga$ and $\ga = \H{\tau}_3 \gr^\dagger \H{\tau}_3$.
Here, $\H{h}$ is a $4\times4$ distribution function, which in systems with spin accumulation can be written
\begin{equation}
  \H{h} = h_0\H{\sigma}_0\H{\tau}_0 + \B{h}_\text{s}\cdot\BH{\sigma}\H{\tau}_3,
\end{equation}
where $h_0$ and $\B{h}_\text{s}$ were introduced earlier.
The $\H{\tau}_n$ are $\H{\sigma}_n$ are Pauli matrices in Nambu and spin space.
As for the retarded component~$\gr$, we analytically use the parametrization~\cite{Eschrig2015a}
\begin{equation}
  \gr = 
  \begin{pmatrix}
    (g_s + \B{g}_t\cdot\B{\sigma}) & (f_s + \B{f}_t\cdot\B{\sigma}) i\sigma_2 \\
    -i\sigma_2 (\tilde{f}_s - \tilde{\B{f}}_t\cdot\B{\sigma}) & -\sigma_2 (\tilde{g}_s - \tilde{\B{g}}_t\cdot\B{\sigma}) \sigma_2 \\
  \end{pmatrix},
\end{equation}
while we numerically use the Riccati parametrization~\cite{SchopohlArxiv}.
General equations for calculating spin supercurrents and spin accumulations from these quasiclassical propagators are derived and presented in the Supplemental {I}nformation.

To determine the propagators above for {\cref{fig:model}}, we have to simultaneously solve the Usadel equation,
\begin{align}
  i\xi^2\nabla(\V{g}\nabla\V{g}) = [\H{\Delta} + \epsilon\H{\tau}_3 ,\, \V{g}]/\Delta_0,
\end{align}
and a selfconsistency equation for the gap~$\Delta$ which {depends on} $\gk$~\cite{Jacobsen2015a}.
The other quantities are the {dirty-limit} coherence length~$\xi$ and bulk gap~$\Delta_0$.
The magnetic insulators in \cref{fig:model} are modelled as spin-active interfaces~\cite{Eschrig2015b, Machon2013, Cottet2009, Cottet2007a}.
{For more details on the numerical model, see the Supplemental Information.}

We assume a fixed distribution function~$\H{h}$, and do not solve any kinetic equation \cite{OuassouArxivA, Bobkova2015a, Silaev2015, BergeretArxiv, Aikebaier2018, Chandrasekhar2008, Belzig1999}. 
This approximation is valid when the superconducting layer is thin compared to its {spin relaxation length.}
{Thus, there is no resistive spin current flowing in the superconductor, as $\nabla \boldsymbol{h}_\text{s}=0$ ensures that there is no gradient in the spin accumulation.}

Finally, we briefly summarize our parameter choices.
The superconductor was taken to have a {length $L = 1.5\xi$}.
The magnetic insulators were described {with $G_\phi/G_\textsc{n} = 0.6$, where $G_\textsc{n}$ is the bulk normal-state conductance of the superconductor, and $G_\phi$ describes the spin-dependent phase-shifts obtained by quasiparticles reflected at a magnetic interface~\cite{Eschrig2015b, Machon2013, Cottet2009, Cottet2007a}.}
Finally, we assumed a constant spin voltage~$V_\text{s}$ throughout the entire superconductor, instead of explicitly modelling the details of the spin source in \cref{fig:model}.
Thus, the junction is treated as a {1D} superconductor with magnetic boundary conditions.
{Our results} are not qualitatively sensitive to these parameter choices.
The main constraints are that superconductivity collapses {if $L/\xi$ is too low and $G_\phi/G_\textsc{n}$ too high, while spin supercurrents become vanishingly small in the opposite limits.}

\prlsection{Numerical results}%
The spin supercurrent in the model considered here is conserved throughout the superconductor~\footnote{We have checked both analytically and numerically that {the spin supercurrent remains conserved in} the presence of spin-flip and spin--orbit impurities, thus extending the equilibrium {results from} Ref.~\cite{Ouassou2017a} to this particular nonequilibrium situation.}.
In fact, the analytical {proof} is straight-forward: the argument in Ref.~\cite{Ouassou2017a} shows that $\nabla\cdot\B{j}_\text{s}^\text{eq} = 0$ as long as $h_0$ is position-independent.
Since the new contribution proposed in this paper $\B{j}_\text{s}^\text{neq} = \B{j}_\text{s}^\text{eq} \times ({i}\B{h}_\text{s}/h_0)$, we conclude that $\nabla\cdot\B{j}_\text{s}^\text{neq} = 0$ for the same physical setups if $\B{h}_\text{s}$ is position-independent.
However, if either $h_0$ or $\B{h}_\text{s}$ becomes inhomogeneous, this argument breaks down, and the spin supercurrent is no longer conserved.

\begin{figure}[t!]
  \includegraphics[width=\columnwidth]{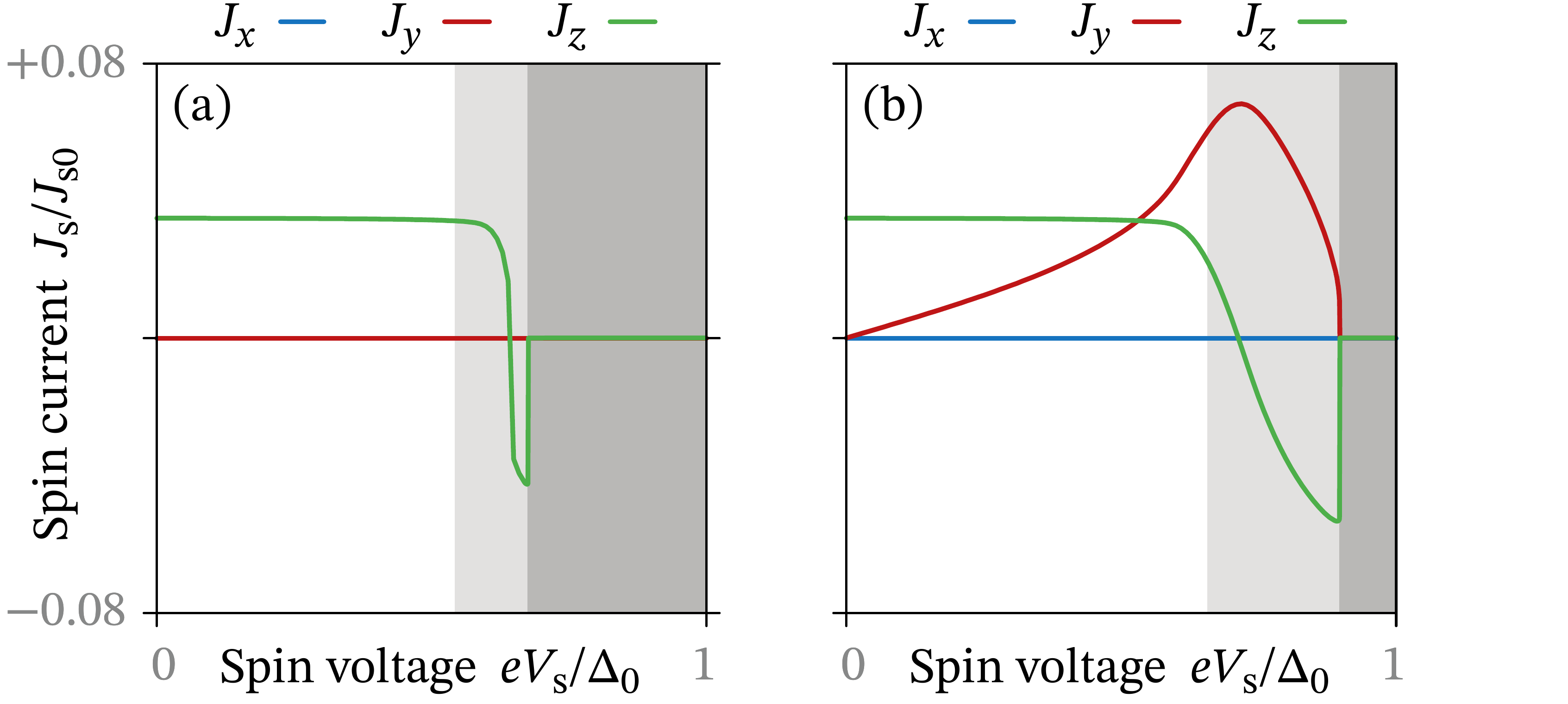}
  \caption%
  {%
    Spin supercurrent $\B{J}_\text{s}$ as a function of spin voltage~$V_\text{s}$.
    This spin voltage corresponds to injected (a) spin-$z$ or (b) spin-$x$ accumulation.
    The light shaded region shows where the system is bistable, and the dark region where superconductivity vanishes.
  }
  \label{fig:voltage}
\end{figure}

In \cref{fig:voltage}, we show the spin supercurrent in the superconductor as a function of spin voltage at a low temperature $T = 0.01T_\text{c}$.
Up until $eV_\text{s} \approx \Delta_0/2$, these results are in perfect agreement with the analytical predictions.
More precisely, we see that a spin-$z$ injection [\cref{fig:voltage}(a)] has no effect on the spin supercurrent, while a spin-$x$ injection [\cref{fig:voltage}(b)] leads to a spin-$y$ supercurrent.
The spin-$y$ supercurrent increases linearly with the spin voltage, again in agreement with the predictions.
Remarkably, the spin-$z$ supercurrent does not decrease as the spin-$y$ supercurrent increases, in contrast to what one {might} intuitively expect.

At low temperatures, we also see that there is a bistable regime at high spin voltages~$eV_\text{s} > \Delta_0/2$.
This means that both a superconducting and normal-state solution exist, which both correspond to local minima in the free energy.
Depending on the dynamics of the system, this can either lead to hysteretic behaviour, or a first-order phase transition.
This first-order phase transition was discussed already in the 1960s by Chandrasekhar and Clogston~\cite{Chandrasekhar1962,Clogston1962}, while the possibility of hysteretic behaviour was suggested much more recently in Refs.~\cite{Bobkova2014,Snyman2009}.
Precisely where in the bistable region the thermodynamic transition point occurs is however difficult to predict within the Usadel formalism, as it is not straight-forward to explicitly evaluate the free energy of each {solution \cite{OuassouArxivA}.}
%{For more information on bistability in superconductors, see e.g.\ Ref.~\cite{OuassouArxivA}.

{Within} the bistable regime, there is a point where the spin-$z$ supercurrent reverses direction as a function of the spin voltage.
{This behaviour can be understood~\cite{OuassouArxivB}~as a spin equivalent of the S/N/S transistor effect~\cite{Wilhelm1998,Baselmans1999} where, according to \cref{eq:energy-dist}, the energy distribution~$h_0$ is also modulated by a spin voltage, and may therefore tune the equilibrium contribution in \cref{eq:j-eq}.}

Since the spin-$y$ supercurrent remains positive for all spin voltages, there exists a point where we get a pure spin-$y$ supercurrent.
In other words, there is a particular spin voltage that causes a $90^\circ$ rotation of the spin supercurrent polarization compared to equilibrium.
The fact that the spin-$y$ supercurrent can remain finite while the spin-$z$ supercurrent goes to zero might at first seem contradictory to our previous explanation $\B{j}_\text{s}^\text{neq} \sim \B{j}_\text{s}^\text{eq} \times ({i}\B{h}_\text{s}/h_0)$.
However, it is the \emph{energy-integrated} spin currents $\B{J}_\text{s} \sim \int\d\epsilon {\im}\left[\B{j}_\text{s}\right]$ that are plotted in \cref{fig:voltage}.
The spin-$y$ current is generated from the \emph{spectral} spin-$z$ current, which remains finite even though the \emph{total} spin-$z$ current is zero.

\begin{figure}[t!]
  \includegraphics[width=\columnwidth]{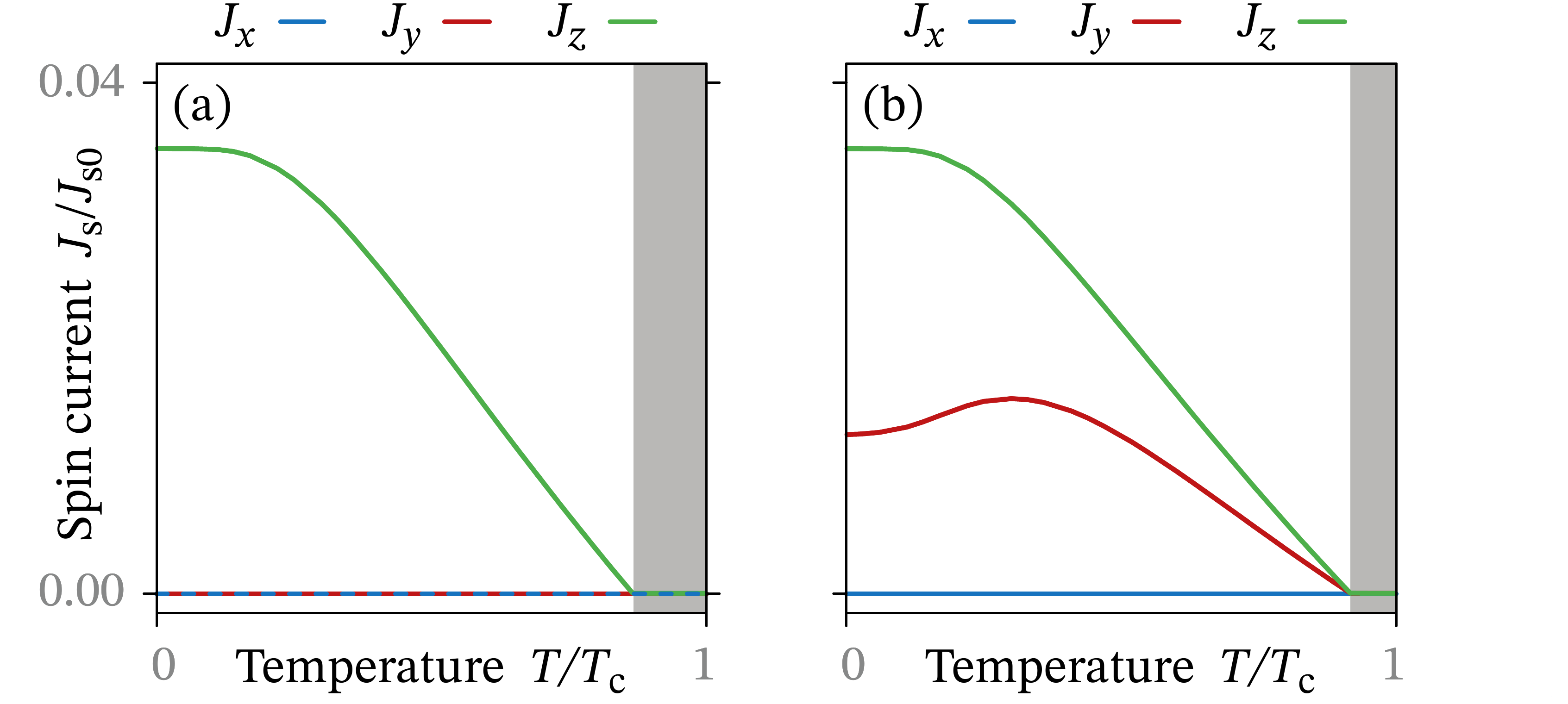}
  \caption%
  {%
    Spin supercurrent $\B{J}_\text{s}$ as a function of temperature~$T$ for a fixed spin voltage~$eV_\text{s} = \Delta_0/4$ in the superconductor.
    This spin voltage corresponds to injected (a) spin-$z$ or (b) spin-$x$ accumulation.
    The shaded region shows where superconductivity vanishes.
  }
  \label{fig:temperature}
\end{figure}

In \cref{fig:temperature}, we show how the spin supercurrent varies as a function of temperature for a fixed spin voltage~$eV_\text{s}=\Delta_0/4$.
Curiously, we find that the spin current increases \emph{linearly} with decreasing temperature in a relatively large parameter regime.
That the spin-$y$ current decreases at the same rate as the spin-$z$ current seems reasonable in light of the equation $\B{j}_\text{s}^\text{neq} \sim \B{j}_\text{s}^\text{eq} \times \B{h}_\text{s}$: if $\B{j}_\text{s}^\text{eq}$ decreases linearly, then $\B{j}_\text{s}^\text{neq}$ should do so as well.
The most important message from \cref{fig:temperature} is perhaps that the nonequilibrium contribution $\B{j}_\text{s}^\text{neq}$ to the spin supercurrent remains significant all the way up to the critical temperature of the junction.
This means that relevant experiments can be performed at any temperature where superconductivity exists.

\prlsection{Discussion}%
In the previous sections, we have shown that injection of a nonequilibrium spin accumulation can be used to generate new spin supercurrent components.
The results are especially encouraging since the nonequilibrium contribution to the spin supercurrent can even be made larger than the equilibrium contribution, and we found that it persists all the way up to the critical temperature of the junction.
Both these features should make it a particularly interesting effect for experimental detection and future device design.
However, there are {some} questions that we have not addressed yet.

The first question is how the spin source in \cref{fig:model} works.
So far, we have simply treated it as a generic device that manipulates the spin distribution~$\B{h}_\text{s}$ inside the superconductor directly.
One alternative is to use a normal metal coupled to a voltage-biased ferromagnet~\cite{Wakamura2014} or half-metal{lic ferromagnet}~\cite{Bobkova2012}.
In that case, the polarization of the magnets enable a charge--spin conversion, thus translating an electric voltage into a spin voltage.
Another possibility would be spin-pumping experiments, where it is a microwave signal that is translated to a spin voltage~\cite{Jeon2018}.
In the limit of weak superconductivity, an expression for the distribution function of a spin-pumped ferromagnet was derived in Ref.~\cite{Houzet2008}.
When the precession frequency~$\Omega$ and cone angle~$\alpha$ are sufficiently small, the result is just a spin voltage $eV_\text{s} = \Omega/2$ along the magnetization direction of the ferromagnet.
This is the relevant limit for the experiment in Ref.~\cite{Jeon2018}: superconductivity inside Ni$_{80}$Fe$_{20}$ is weak, $\Omega \approx 0.5$~meV is much smaller than its magnetic exchange field, and $\alpha \approx 1^\circ$ should be small enough to use the leading-order expansions $\sin \alpha \approx \alpha$ and $\cos \alpha \approx 1$.

In all these cases, the spin source necessarily contains magnetic elements, and one challenge would be how to prevent the spin source from affecting the \emph{equilibrium} spin current.
One solution might be to embrace the existing magnets in \cref{fig:model}: one could use the same magnets to generate the equilibrium spin supercurrent and for spin injection.
This spin injection may them be performed either using spin pumping---or if the magnets are sufficiently thin for electron tunneling---by placing voltage-biased contacts on top of the magnets.
One complication with this strategy is that since the resulting spin accumulation will necessarily be inhomogeneous, both spin supercurrents and resistive spin currents have to coexist.

%Another question {relates to how} one might measure these spin supercurrents.
{How to} {directly measure} a spin supercurrent is {an open question}, although suggestions have {recently been proposed}~\cite{RisinggardArxiv}.
Indirect measurements of spin supercurrents, on the other hand, have already been performed experimentally.
Most of these rely on measuring dissipationless charge currents through strongly polarized materials~\cite{Keizer2006b,Anwar2010,Robinson2010,Robinson2010a,Khaire2010a,Witt2012,Robinson2014,Egilmez2014}.
Since only $\ket{\up\up}$ and $\ket{\dn\dn}$ pairs can penetrate over longer distances, and the polarization breaks the degeneracy between them, one can infer the existence of spin supercurrents from the measured charge supercurrents.

One solution to the measurement problem might be to look for an \emph{inverse} effect.
We have shown that spin injection into a superconductor results in a torque on the spins transported by the equilibrium spin supercurrent.
However, this interaction should cause a reaction torque {on} the spin source, which might be possible to detect.
For instance, in a setup similar to Ref.~\cite{Wakamura2014}, this reaction torque might directly affect the nonlocal spin conductance.
Similarly, in a spin-pumping setup, this might affect the FMR linewidths.
In both cases, this reaction torque should only exist when there is an equilibrium spin supercurrent $\B{j}_\text{s}^\text{eq} \sim \B{m}_{{1}}\times\B{m}_{{2}}$ to interact with, so it should depend on the magnetic configuration of the device.

\prlsection{Conclusion}%
We have shown analytically and numerically that if a system harbors a spin supercurrent~$\B{j}_\text{s}^\text{eq}$ in equilibrium, then a spin injection~$\B{h}_\text{s}$ creates a new component $\B{j}_\text{s}^\text{neq} \sim \B{j}_\text{s}^\text{eq} \times \B{h}_\text{s}$.
This effect can be intuitively understood as the injected spins exerting a torque on the spins transported by the equilibrium spin supercurrent, generating a component that is perpendicular to both.
These results have implications for the control of spin supercurrents in novel superconducting spintronics devices.

%-------------------------------------------------------------------------------%
%                                ACKNOWLEDGEMENTS                               %
%-------------------------------------------------------------------------------%

\begin{acknowledgments}
  We thank M.~Amundsen and V.~Risinggård for useful discussions.
  The numerics {were} performed on resources provided by UNINETT Sigma2---the {N}ational infrastructure for high performance computing and data storage in Norway.
  {J.A.O. and J.L. were} supported by the Research Council of Norway through grant 240806, and through its Centres of Excellence funding scheme grant 262633 ``\emph{QuSpin}''.
  {J.W.A.R. acknowledges the EPSRC--JSPS ``\textit{OSS SuperSpin}'' International Network Grant (EP/P026311/1) and Programme Grant (EP/N017242/1).}
\end{acknowledgments}

%-------------------------------------------------------------------------------%
%                                  BIBLIOGRAPHY                                 %
%-------------------------------------------------------------------------------%

%\clearpage

% Bibtex bibliography
%\bibliography{references}

% Bibliography preamble
%
\end{document}

\title{Supplemental information}
\author{Jabir Ali Ouassou}
\affiliation{Center for Quantum Spintronics, Department of Physics, Norwegian \\ University of Science and Technology, NO-7491 Trondheim, Norway}
\author{Jason W. A. Robinson}
\affiliation{Department of Materials Science and Metallurgy, University of Cambridge, \\ 27 Charles Babbage Road, Cambridge CB3 0FS, {United Kingdom}}
\author{Jacob Linder}
\affiliation{Center for Quantum Spintronics, Department of Physics, Norwegian \\ University of Science and Technology, NO-7491 Trondheim, Norway}
 
\begin{abstract}
  \noindent
  \cref{sec:transport} provides a self-contained derivation of the charge and spin transport equations.
  Specifically, we start with the definitions of particle densities from quantum field theory, and derive quasiparticle continuity equations.
  The results are used to derive quasiclassical results for the charge accumulation, spin accumulation, charge current, and spin current.
  \cref{sec:current-eq} then uses these results to derive an analytical expression for the spin supercurrent in materials with spin accumulation.
  The result is used to explain the predictions in the main manuscript.
  {Finally, in \cref{sec:numerics}, we discuss some technical details about the numerical model used for Figs.~2--3 in the main article.}
\end{abstract}

\maketitle

\section{Charge and spin transport}\label{sec:transport}
\subsection{Quasiparticle accumulations}
There are four relevant species of quasiparticles in the systems that we will consider: namely electrons and holes, which each have two distinct spin projections.
These have the densities
\begin{align}
  \label{def:quasiparticle-density}
  n_{e\up} (\B r, t) &\coloneq \bexpect{ \create[\up]  (\B r, t) \, \destroy[\up] (\B r, t) }, \\
  n_{e\dn} (\B r, t) &\coloneq \bexpect{ \create[\dn]  (\B r, t) \, \destroy[\dn] (\B r, t) }, \\
  n_{h\up} (\B r, t) &\coloneq \bexpect{ \destroy[\up] (\B r, t) \, \create[\up]  (\B r, t) }, \\
  n_{h\dn} (\B r, t) &\coloneq \bexpect{ \destroy[\dn] (\B r, t) \, \create[\dn]  (\B r, t) },
\end{align}
where $\create[\sigma]$ and $\destroy[\sigma]$ are standard creation and annihilation operators.
For comparison, the propagators are defined as~\cite{Rammer1986,Belzig1999,Chandrasekhar2008}:
\begin{align}
  \!\! \GR_{\sigma\sigma'}(\B r, t; \B r'\!, t') &\coloneq -i\expect{ \anticomm{ \destroy[\sigma](\B r, t) }{ \create[\sigma'](\B r'\!, t') } }\, \theta(t - t'), \\
  \!\! \GA_{\sigma\sigma'}(\B r, t; \B r'\!, t') &\coloneq +i\expect{ \anticomm{ \destroy[\sigma](\B r, t) }{ \create[\sigma'](\B r'\!, t') } }\, \theta(t' - t), \\
  \!\! \GK_{\sigma\sigma'}(\B r, t; \B r'\!, t') &\coloneq -i\expect{     \comm{ \destroy[\sigma](\B r, t) }{ \create[\sigma'](\B r'\!, t') } }, \label{def:keldysh-propagator}
\end{align}
where the subscripts $\sigma$ and $\sigma'$ denote possible spin projections.
Combining these definitions, we see that the quasiparticle densities are directly related to the equal-coordinate propagators:
\begin{align}
  n_{e\sigma} &= \frac{i}{2} 
                  [ \GR_{\sigma\sigma} 
                  - \GA_{\sigma\sigma} 
                  - \GK_{\sigma\sigma} ], \\
  n_{h\sigma} &= \frac{i}{2} 
                  [ \GR_{\sigma\sigma} 
                  - \GA_{\sigma\sigma} 
                  + \GK_{\sigma\sigma} ].
\end{align}
These expressions can be used to calculate the spin-resolved density of electrons and holes, respectively.
Note that holes carry both opposite charge and opposite spin compared to electrons~%
\footnote{The fact that holes carry opposite spins from electrons can be seen from the field operators.
In our formalism, a ``spin-$\sigma$ hole'' refers to the quasiparticles counted by $\expect{ \destroy[\sigma] \create[\sigma] }$.
This means that the creation operator for spin-up holes is the annihilation operator for spin-up electrons.
Annihilating a spin-up electron \emph{reduces} the spin in the system by $\hbar/2$, so a spin-up hole must actually {carry net spin in the down-direction}.
In other words, a ``spin-up hole'' is ``a hole that is left when a spin-up electron is removed'', not ``a hole that physically carries spin-up''.}.
The charge and spin accumulations are then found by multiplying each quasiparticle density with their respective charges or spins, and summing up their contributions:
\begin{align}
  \label{def:density-charge-spin}
  \rho_\text{e} &\coloneq e\,             {\frac{1}{2}} [ +n_{e\up} + n_{e\dn} - n_{h\up} - n_{h\dn} ], \\
  \rho_z &\coloneq \frac{\hbar}{2} {\frac{1}{2}} [ +n_{e\up} - n_{e\dn} - n_{h\up} + n_{h\dn} ],
\end{align}
where we use the convention that $e$ is the \emph{electron charge}~($e<0$).
{The prefactors $1/2$ are required to prevent double-counting, and can be explained as follows.
If we add one physical electron to the system, then the charge of the system increases by~$e$.
However, the number of electrons increases by one, and the number of holes decreases by one, meaning that the difference between electrons and holes increases by \emph{two}.
Thus, when the charge density~$\rho_\text{e}$ is described in terms of \emph{both} electrons and holes, we need an extra factor~$1/2$ to get the right physical charge.
The same logic applies to the spin density~$\rho_z$.}
We can rewrite the results in terms of the propagators above, and recognize the remaining sum as a trace over spins:
\begin{align}
  \rho_\text{e} &= -{\frac{i}{2}}e\,\tr[\sigma_0 \GK],\\
  \rho_z &= -{\frac{i}{2}}\frac{\hbar}{2} \tr[\sigma_3 \GK].
\end{align}
There is nothing special about the spin-$z$ axis, so it is straight-forward to generalize this result to arbitrary spin-projections:
\begin{align}
  \rho_\text{e}   &= -{\frac{i}{2}}e\,\tr[\sigma_0 \GK],\\
  \B\rho_\text{s} &= -{\frac{i}{2}}\frac{\hbar}{2} \tr[\B\sigma \GK],
\end{align}
where $\B{\sigma} = (\sigma_1,\sigma_2,\sigma_3)$ is the Pauli vector.
From the definition of the Keldysh propagator above, we can also use the identity $\cc{\expect{AB}} = \expect{\hc{B}\!\hc{A\!}}$ to show that $\GKS_{\sigma\sigma} = -\GK_{\sigma\sigma}$.
This means that $\GK_{\sigma\sigma}$ is imaginary, which makes $\rho_\text{e}, \rho_\text{s} \sim i\GK$ manifestly real.
For later convenience, we will therefore write this out explicitly:
\begin{align}
  \rho_\text{e} &= {-\frac{1}{2}}e\,\re\tr[i\sigma_0 \GK],\\
    \B\rho_\text{s} &= {-\frac{1}{2}}\frac{\hbar}{2} \re\tr[i\B\sigma \GK].
\end{align}

\subsection{Quasiparticle currents}
Now that we know the charge and spin accumulations, the next step is to find the corresponding currents.
To derive these, we go back to the quasiparticle densities defined in \cref{def:quasiparticle-density}:
\begin{align}
  n_{e\sigma} (\B r, t) &\coloneq \bexpect{ \create[\sigma]  (\B r, t) \, \destroy[\sigma] (\B r, t) }, \\
  n_{h\sigma} (\B r, t) &\coloneq \bexpect{ \destroy[\sigma] (\B r, t) \, \create[\sigma]  (\B r, t) }. 
\end{align}
To rigorously derive expressions for the charge and spin currents, we will use the definitions above to look for quasiparticle continuity equations on the form
\begin{equation}
  \label{eq:qp-continuity}
  \partial_t n_{\tau\sigma} + \nabla\cdot\B{j}_{\tau\sigma} = q_{\tau\sigma},
\end{equation}
where $\B{j}_{\tau\sigma}$ is the particle- and spin-resolved current density we are interested in, while $q_{\tau\sigma}$ represents possible source terms.

{We start by differentiating the densities with respect to time:}
\begin{align}
  \partial_t n_{e\sigma} &= \bexpect{ (\partial_t\! \create[\sigma] ) \, \destroy[\sigma] + \create[\sigma]  \, (\partial_t\! \destroy[\sigma]) }, \\
  \partial_t n_{h\sigma} &= \bexpect{ (\partial_t\! \destroy[\sigma]) \, \create[\sigma]  + \destroy[\sigma] \, (\partial_t\! \create[\sigma] ) }. 
\end{align}
We can rewrite the above using the Heisenberg equation of motion for the field operators.
{Note that any contributions to the continuity equation arising from non-derivative terms in the Hamiltonian---such as a superconducting gap or an exchange field---can be incorporated into the source term~$q$.
Thus, for the purposes of deriving current equations, it is sufficient to consider only derivative terms.}
{Whether or not the currents we derive are \emph{conserved} currents can be checked at the end of the derivation, by substituting the Usadel equation into the final quasiclassical current equations~\cite{Ouassou2017a,Jacobsen2016}.}
If we for simplicity disregard gauge fields for now, the equations reduce to:
\begin{align}
  \partial_t \destroy[\sigma] &= +\frac{i}{2m} \nabla^2 \destroy[\sigma], \\
  \partial_t \create[\sigma]  &= -\frac{i}{2m} \nabla^2 \create[\sigma].
\end{align}
We then substitute these back into the equations for $\partial_t n_{\tau\sigma}$:
\begin{align}
  \partial_t n_{e\sigma} &= -\frac{i}{2m} \bexpect{ (\nabla^2 \create[\sigma] ) \, \destroy[\sigma] - \create[\sigma]  \, (\nabla^2 \destroy[\sigma]) }, \\
  \partial_t n_{h\sigma} &= +\frac{i}{2m} \bexpect{ (\nabla^2 \destroy[\sigma]) \, \create[\sigma]  - \destroy[\sigma] \, (\nabla^2 \create[\sigma] ) }. 
\end{align}
Thanks to cancellation of cross-terms, these can be factorized:
\begin{align}
  \partial_t n_{e\sigma} &= -\frac{i}{2m} \nabla \cdot \bexpect{ (\nabla \create[\sigma] ) \, \destroy[\sigma] - \create[\sigma]  \, (\nabla \destroy[\sigma]) }, \\
  \partial_t n_{h\sigma} &= +\frac{i}{2m} \nabla \cdot \bexpect{ (\nabla \destroy[\sigma]) \, \create[\sigma]  - \destroy[\sigma] \, (\nabla \create[\sigma] ) }. 
\end{align}
Comparing this to \cref{eq:qp-continuity}, we conclude that:
\begin{align}
  j_{e\sigma} &= -\frac{i}{2m} \bexpect{ (\nabla \create[\sigma] ) \, \destroy[\sigma] - \create[\sigma]  \, (\nabla \destroy[\sigma]) }, \\
  j_{h\sigma} &= +\frac{i}{2m} \bexpect{ (\nabla \destroy[\sigma]) \, \create[\sigma]  - \destroy[\sigma] \, (\nabla \create[\sigma] ) }. 
\end{align}
As a mathematical trick, let us now use different coordinates $\destroy[\sigma] = \destroy[\sigma](\B r,t)$ and $\create[\sigma] = \create[\sigma](\B r', t')$ for the field operators, where we let $\B r' \rightarrow \B r$ and $t' \rightarrow t$ in the end.
In this case, the differential operators acting on the field operators can be factored out of the expectation value without ambiguity:
\begin{align}
  \label{eq:qp-current-density}
  j_{e\sigma} &= -\frac{i}{2m} (\nabla' - \nabla) \bexpect{ \create[\sigma] (\B r',t')\, \destroy[\sigma](\B r,  t ) }, \\
  j_{h\sigma} &= -\frac{i}{2m} (\nabla' - \nabla) \bexpect{ \destroy[\sigma](\B r, t )\, \create[\sigma] (\B r', t') }. 
\end{align}
We are now ready to define the charge and spin current densities.
In correspondence with \cref{def:density-charge-spin}, we define these as:
\begin{align}
  \label{def:current-density-1}
  J_\text{e} &\coloneq e\,             {\frac12} [+j_{e\up} +j_{e\dn} -j_{h\up} -j_{h\dn}],\\
  \label{def:current-density-2}
  J_{z} &\coloneq \frac{\hbar}{2} {\frac12} [+j_{e\up} -j_{e\dn} -j_{h\up} +j_{h\dn}].
\end{align}
Substituting \cref{eq:qp-current-density} into \cref{def:current-density-1,def:current-density-2}, comparing the results to the Keldysh propagator in \cref{def:keldysh-propagator}, and recognizing the results as {traces in spin space, we conclude:}
\begin{align}
  J_\text{e} &= -e\,\frac{1}{{4}m}              (\nabla' - \nabla)\, \tr[\sigma_0 \GK], \\
  J_{z} &= -\frac{\hbar}{2} \frac{1}{{4}m} (\nabla' - \nabla)\, \tr[\sigma_3 \GK].
\end{align}
Generalizing to all spin projections, we obtain the final results:
\begin{align}
  J_\text{e}     &= -e\,\frac{1}{{4}m}              (\nabla' - \nabla)\, \tr[\sigma_0   \GK], \\
  \B{J}_\text{s} &= -\frac{\hbar}{2} \frac{1}{{4}m} (\nabla' - \nabla)\, \tr[\B{\sigma} \GK].
\end{align}

We wish to point out that these currents are manifestly real.
From the definition of the Keldysh propagator, we see that:
\begin{equation}
  (\nabla' - \nabla) [\GKS(\B r, t; \B r', t')] = (\nabla' - \nabla) [-\GK(\B r', t'; \B r, t)].
\end{equation}
But which set of coordinates we chose to call $(\B r,t)$ and $(\B r',t')$ was arbitrary, and should not affect the physical results, since we are considering the limit $\B r', t' \rightarrow \B r,t$ anyway.
This means that we can interchange the coordinates $(\B r, t)$ and $(\B r',t')$ on the right-hand side of the equation, \emph{as long as we do this consistently for every factor simultaneously}.
The coordinate interchange leads to a sign flip in $(\nabla' - \nabla)$ which cancels the minus sign inside the brackets, and makes the two sides of the equation equal.
This lets us conclude that $(\nabla' - \nabla) \GKS = (\nabla' - \nabla) \GK$, which in turn implies that the charge and spin currents are real.
For later convenience, we can therefore rewrite the above as
\begin{align}
  J_\text{e}     &= -e\, \frac{1}{{4}m} \re\tr[\sigma_0 (\nabla' - \nabla) \, \GK], \\
  \B{J}_\text{s} &= -\frac{\hbar}{2} \frac{1}{{4}m} \re\tr[\B{\sigma} (\nabla' - \nabla) \, \GK].
\end{align}

\subsection{Quasiclassical and diffusive limits}
To derive equations we can use together with the Usadel equation, we now follow the standard prescription for {taking} the quasiclassical and diffusive limits~\cite{Kopnin2001,Belzig1999,Chandrasekhar2008,Rammer1986}.
The net change to the Keldysh propagator and its derivative are then:
\begin{align}
  i\GK &\rightarrow \frac{1}{4} N_0\! \int\limits_{-\infty}^{+\infty} \d\epsilon\, \expect{ g^{\text{\smaller K}} }_{\text{\smaller F}}^{\vphantom{*}}, \\
  \frac{1}{2m} (\nabla' - \nabla) \GK &\rightarrow \frac{1}{4} N_0\! \int\limits_{-\infty}^{+\infty} \d\epsilon\, \expect{ \B{v} g^{\text{\smaller K}} }_{\text{\smaller F}}^{\vphantom{*}},
\end{align}
where $\B{v} \coloneq \B{p}/m$ is interpreted as the quasiparticle velocity, $N_0$ is the density of states at the Fermi level, and $\expect{ \cdots }_{\text{\smaller F}}^{\vphantom{*}}$ refers to the average over the Fermi surface.
From the derivation of the Usadel equation, we also know that in the diffusive limit the Fermi-surface averages can be written~\cite{Kopnin2001,Tokatly2017}
\begin{align}
  \expect{\V{g}}_{\text{\smaller F}}^{\vphantom{*}} &\coloneq \V{g}_s, & 
  \expect{\B{v} \V{g}}_{\text{\smaller F}}^{\vphantom{*}} &\approx -D (\V{g}_s \tilde\nabla \V{g}_s),
\end{align}
where $\tilde\nabla$ is a {gauge-covariant derivative including the electromagnetic vector potential and spin-orbit interactions}~\cite{Tokatly2017,Jacobsen2015a,Kopnin2001}, $\V{g}_s$ is the isotropic propagator, and $D$ is the diffusion constant.
We drop the subscripts on the isotropic propagators $\V{g}_s$, and substitute the above into the accumulations and currents:
\begin{align}
  \rho_\text{e}    &= -e\,             \frac{1}{{8}} N_0 \!\int\limits_{-\infty}^{+\infty} \d\epsilon\, \re\tr[\sigma_0   g^{\text{\smaller K}}], \\
  \B\rho_\text{s}  &= -\frac{\hbar}{2} \frac{1}{{8}} N_0 \!\int\limits_{-\infty}^{+\infty} \d\epsilon\, \re\tr[\B\sigma   g^{\text{\smaller K}}], \\
  J_\text{e}     &= +e\,             \frac{1}{{8}} N_0 \!\int\limits_{-\infty}^{+\infty} \d\epsilon\, \re\tr[\sigma_0   \B{I}^{\text{\smaller K}}], \\
  \B{J}_\text{s} &= +\frac{\hbar}{2} \frac{1}{{8}} N_0 \!\int\limits_{-\infty}^{+\infty} \d\epsilon\, \re\tr[\B{\sigma} \B{I}^{\text{\smaller K}}],
\end{align}
where we have reintroduced the matrix current $\V{\B I} \coloneq D(\V{g}\tilde\nabla \V{g})$.
Note that these equations only depend on the ``electronic'' part of the propagators in Nambu space, which in reality contains information about both the electrons and holes in the system.

All these results can be written as integrals over only positive energies using the symmetries of the Nambu-space matrices
\begin{align}
  \H\sigma_n\K{g}(\epsilon) &= 
    \begin{pmatrix}
      +\sigma_n   g^{\text{\smaller K} }(+\epsilon) & +\sigma_n f^{\text{\smaller K} }(+\epsilon) \\
      +\sigma_n^* {f}^{\text{\smaller K}*}(-\epsilon) & +\sigma_n^* g^{\text{\smaller K}*}(-\epsilon)
    \end{pmatrix},
    \\
  \H\sigma_n \K{\B I}(\epsilon) &= 
    \begin{pmatrix}
      +\sigma_n   \B{I}^{\text{\smaller K} }(+\epsilon) & +\sigma_n   {\B{J}}^{\text{\smaller K} }(+\epsilon) \\
      -\sigma_n^*{\B{J}}^{\text{\smaller K}*}(-\epsilon) & -\sigma_n^* \B{I}^{\text{\smaller K}*}(-\epsilon)
    \end{pmatrix}.
\end{align}
In other words, the negative-energy contributions can be recast in terms of the lower-right blocks; and since take the real part of the results, the complex conjugations are irrelevant.
The remaining structure can be recognized as a trace over Nambu space, yielding the final quasiclassical transport equations
\begin{align}
  \rho_\text{e}    &= -e\,             \frac{1}{{8}} N_0 \!\int\limits_{0}^{\infty} \d\epsilon\, \re\tr[\H{\tau}_0 \H{\sigma}_0   \K g], \\
  \B\rho_\text{s}  &= -\frac{\hbar}{2} \frac{1}{{8}} N_0 \!\int\limits_{0}^{\infty} \d\epsilon\, \re\tr[\H{\tau}_0 \H{\B\sigma}   \K g], \label{eq:spin-acc-final} \\
  J_\text{e}     &= +e\,             \frac{1}{{8}} N_0 \!\int\limits_{0}^{\infty} \d\epsilon\, \re\tr[\H{\tau}_3 \H{\sigma}_0   \K{\B I}], \\
  \B{J}_\text{s} &= +\frac{\hbar}{2} \frac{1}{{8}} N_0 \!\int\limits_{0}^{\infty} \d\epsilon\, \re\tr[\H{\tau}_3 \H{\B{\sigma}} \K{\B I}].
\end{align}
Note that $\H{\B{\sigma}} \K{\B I}$ should be interpreted as an outer product between two vectors, which results in a rank-2 tensor.
This is because a general description of spin transport requires both a direction of transport $\sim\!\H{\B{I}}^{\text{\smaller K}}$ and a spin orientation $\sim\!\H{\B{\sigma}}$.
\vspace{0.01ex}

\subsection{Higher-order gauge contributions}
{The equations of motion for the field operators also include first-order derivative terms in systems with electromagnetic \cite{Kopnin2001,Morten2005} or spin-orbit~\cite{Jacobsen2015a,Bergeret2013a,Bergeret2014a,Tokatly2017} gauge fields.}
{If we ignore} all other terms in the Hamiltonian, {these derivative terms} give the following Heisenberg equations: 
\begin{align}
  \partial_t \destroy[\sigma] &= \frac{1}{m} \B{A}_{\sigma\sigma'} \cdot (\nabla \destroy[\sigma']), \\
  \partial_t \create[\sigma]  &= \frac{1}{m} (\nabla \create[\sigma']) \cdot \B{A}_{\sigma'\sigma},
\end{align}
where we implicitly sum over the spin index~$\sigma'$.
Going through the same kind of derivations as without the gauge fields, we find that we basically just have to make the following replacement in the results right before taking the quasiclassical limit:
\begin{equation}
  \B{p} \GK \rightarrow \frac{1}{2}\anticomm{ \GK\! }{ \B{p} - \B{A} }.
\end{equation}
Note that the gauge fields also affects charge and spin transport in a different way, since they also appear as covariant derivatives~$\tilde{\nabla}(\,\cdot\,) = \nabla(\,\cdot\,) - i[\B{A},\,\cdot\,]$ in the matrix current~$\V{I} = D\V{g}\tilde{\nabla}\V{g}$.

\section{Nonequilibrium supercurrents}\label{sec:current-eq}
\subsection{Supercurrents vs. resistive currents}
As shown in previous sections, the total spin current~$\B{J}_\text{s}$ can in the quasiclassical limit be calculated as an energy integral,
\begin{equation}
  \B{J}_\text{s} = \frac{\hbar}{2} N_0 \int_0^\infty \d\epsilon\, \B{j}_\text{s},
\end{equation}
where the spectral spin current~$\B{j}_\text{s} \coloneq \re\tr[\hat{\tau}_3 \hat{\B{\sigma}} \hat{\B{I}}^{\text{\smaller K}}]{/8}$ and the matrix current $\hat{\B{I}} \coloneq D\V{g}{\nabla}\V{g}$.
If we substitute the parametrization $\K{g} = \R{g}\H{h} - \H{h}\A{g}$ into the definition of the matrix current, we find that its Keldysh component can be expanded as
\begin{equation}
  \begin{aligned}
    \BK{I} = \,&D[(\R{g}\nabla\R{g}) \H{h} - \H{h} (\A{g}\nabla\A{g})] \\
           + \,&D[(\nabla\H{h}) - \R{g} (\nabla\H{h}) \A{g}].
  \end{aligned}
  \label{eq:mat-current-decom}
\end{equation}
The terms on the first line may be finite even for a homogeneous distribution function~$\H{h}$, and produces spin currents even in equilibrium.
Furthermore, they are sensitive to the phase-winding of the superconducting condensate via $\R{g}\nabla\R{g}$ and $\A{g}\nabla\A{g}$.
We therefore identify this as a supercurrent contribution.
The terms on the second line, however, are proportional to $\nabla \H{h}$.
This current contribution both requires an inhomogeneous distribution function, and is insensitive to the phase-winding of the superconducting condensate, and has to be a resistive current.

{In this work, we are primarily interested in generating a spin supercurrent from a nonequilibrium spin accumulation.
We therefore limit our attention to systems with a position-independent distribution function~$\gh$ that has an excited spin mode.
Since we assume $\nabla\gh = 0$, the second line of \cref{eq:mat-current-decom} disappears, and only the supercurrent contribution remains:
\begin{equation}
  \B{j}_\text{s} = \frac{1}{{8}} \re \tr \big[ \gh \nambu 3 \hat{\B{\sigma}} (\gr\nabla\gr) - \nambu 3 \hat{\B{\sigma}} \gh(\gr^\dagger \nabla \gr^\dagger) \big].
  \label{eq:supercurrent}
\end{equation}
As for the distribution function, it can be parametrized as}
\begin{equation}
  \gh = h_0 \spin 0 \nambu 0  + \B{h}_\text{s} \cdot \H{\B{\sigma}} \nambu 3 ,
  \label{eq:distribution-ansatz}
\end{equation}
where $\B{h}_\text{s}$ points along the net quantization axis of the accumulated spins, and the magnitudes of the modes above are
\begin{align}
  h_0(\epsilon) &= \tfrac12 \tanh[(\epsilon+eV_\text{s})/2T] + \tfrac12 \tanh[(\epsilon-eV_\text{s})/2T], \\
  h_\text{s}(\epsilon) &= \tfrac12 \tanh[(\epsilon+eV_\text{s})/2T] - \tfrac12 \tanh[(\epsilon-eV_\text{s})/2T].
\end{align}
Note that the energy mode~$h_0$ and spin mode~$h_\text{s}$ are odd and even functions of energy, respectively.
{We have parametrized the spin mode in terms of a spin voltage~$V_\text{s} \coloneq (V_\up - V_\dn)/2$, where $V_\sigma$ are the effective potentials experienced by spin-$\sigma$ quasiparticles \cite{OuassouArxivA,BergeretArxiv,Bauer2012}.
The spin mode~$\B{h}_\text{s}$ is related to the spin accumulation in \cref{eq:spin-acc-final} by an energy integral $\B{\rho}_\text{s} \sim \int \d\epsilon\, N(\epsilon)\,\B{h}_\text{s}(\epsilon)$, where $N(\epsilon)$ is the density of states~\cite{BergeretArxiv}.}

\subsection{Expansion in Pauli matrices}
Once we substitute \cref{eq:distribution-ansatz} into \cref{eq:supercurrent}, there are a few subtleties to be careful about.
To handle these, without yet introducing all the details of the singlet/triplet-decomposition, {we first expand} $\gr\nabla\gr$ directly in terms of Pauli matrices:
\begin{equation}
  \gr\nabla\gr \coloneq \B{\alpha}\cdot\spinv\nambu3 + \B{\beta}\cdot\spinv\nambu0 + \gamma\,\spin0\nambu3 + \delta\,\spin0\nambu0 + \hat{\epsilon}.
  \label{eq:trace-decomposition}
\end{equation}
The first four terms parametrizes a general block-diagonal matrix, while the last term $\hat{\epsilon}$ represents off-block-diagonal parts.
Since the distribution~$\hat{h}$ can always be chosen to be block-diagonal, $\hat{\epsilon}$ does not contribute to the trace in \cref{eq:supercurrent}.
The other coefficients are found by taking appropriate traces:
\begin{equation}
  \begin{aligned}
    \B{\alpha} &= \frac14 \tr[\spinv\nambu3 \gr\nabla\gr], &
    \B{\beta} &= \frac14 \tr[\spinv\nambu0 \gr\nabla\gr], \\
       \gamma  &= \frac14 \tr[\spin0\nambu3 \gr\nabla\gr], &
       \delta  &= \frac14 \tr[\spin0\nambu0 \gr\nabla\gr].
  \end{aligned}
  \label{eq:trace-coefficients}
\end{equation}
We parametrize $\gr^\dagger \nabla \gr^\dagger$ using coefficients $\underline{\B{\alpha}}, \underline{\B{\beta}}, \underline{\gamma}, \underline{\delta}$ that are defined in the same manner as above.

{We} will now argue that the parameter~$\delta$ is identically zero.
{By differentiating the normalization condition $(\gr)^2 = 1$, one can show that the retarded propagator anticommutes with its gradient, $\{\gr, \nabla\gr\} = 0$.}
This {identity} can be rewritten
\begin{equation}
  \gr(\nabla\gr) = -(\nabla\gr)\gr.
  \label{eq:anticomm}
\end{equation}
Let us now trace both sides of the equation, and use the cyclic rule $\tr[\hat A \hat B] = \tr[\hat B \hat A]$ on the right-hand side,
\begin{equation}
  \tr[\gr(\nabla\gr)] = -\tr[\gr(\nabla\gr)].
\end{equation}
Since $\spin0\nambu0$ is an identity matrix, we see from \cref{eq:trace-coefficients} that:
\begin{equation}
  \delta = -\delta.
\end{equation}
In other words, $\delta = 0$ is always satisfied, as any other conclusion would violate the normalization condition~$(\gr)^2 = 1$.

{Next, to clarify another subtlety,} we need to derive some trace identities.
By explicitly writing out the matrix products and using $\spinv = \text{diag}(\B{\sigma}, \B{\sigma}^*)$, one can show that
\begin{align}
  \tr[(\B{a}\cdot\spinv)(\B{b}\cdot\spinv)\spinv\nambu0] 
  &=   \tr[(\B{a}\cdot\B{\sigma}^{\phantom{*}})(\B{b}\cdot\B{\sigma}^{\phantom{*}})\B{\sigma}^{\phantom{*}}] \notag \\
  &\,+ \tr[(\B{a}\cdot\B{\sigma}^*)(\B{b}\cdot\B{\sigma}^*)\B{\sigma}^*], \\[0.5ex]
  \tr[(\B{a}\cdot\spinv)(\B{b}\cdot\spinv)\spinv\nambu3] 
  &=   \tr[(\B{a}\cdot\B{\sigma}^{\phantom{*}})(\B{b}\cdot\B{\sigma}^{\phantom{*}})\B{\sigma}^{\phantom{*}}] \notag \\
  &\,- \tr[(\B{a}\cdot\B{\sigma}^*)(\B{b}\cdot\B{\sigma}^*)\B{\sigma}^*].
\end{align}
Products of spin matrices in general satisfy ${(\B{a}\cdot\B{\sigma})(\B{b}\cdot\B{\sigma})} = {(\B{a}\cdot\B{b})} + {i(\B{a}\times\B{b})\cdot\B{\sigma}}$;
multiplying by $\B{\sigma}$ and taking the trace, we find the associated trace rule ${\tr[(\B{a}\cdot\B{\sigma})(\B{b}\cdot\B{\sigma})\B{\sigma}]} = {+2i(\B{a}\times\B{b})}$.
However, if we complex-conjugate before taking the trace, we uncover another identity ${\tr[(\B{a}\cdot\B{\sigma}^*)(\B{b}\cdot\B{\sigma}^*)\B{\sigma}^*]} = {-2i(\B{a}\times\B{b})}$.
A geometric motivation for the sign difference is that if the basis $\B{\sigma} = (\sigma_1, \sigma_2, \sigma_3)$ defines a right-handed coordinate system, then $\B{\sigma}^* = (\sigma_1, -\sigma_2, \sigma_3)$ has to define a left-handed one---and this inverts the right-hand rule that cross-products usually satisfy.
With the aid of the results above, we see that
\begin{align}
  \tr[(\B{a}\cdot\spinv)(\B{b}\cdot\spinv)\spinv\nambu0] 
  &=   0, \\
  \tr[(\B{a}\cdot\spinv)(\B{b}\cdot\spinv)\spinv\nambu3] 
  &=  4i(\B{a}\times\B{b}).
\end{align}
This is the subtle trap alluded to above: due to the way we define~$\spinv = \text{diag}(\B{\sigma},\B{\sigma}^*)$, the generalization of the Pauli cross-product identity to matrices in Nambu space requires an extra factor~$\nambu3$ in the trace to produce a nonzero result.

We now substitute \cref{eq:trace-decomposition,eq:distribution-ansatz} into \cref{eq:supercurrent}.
With the identities above, we see that the only contributions are:
\begin{equation}
  \B{j}_\text{s} = {\tfrac12} h_0     \re \big[ \B{\alpha} - \underline{\B{\alpha}} \, \big]
  + {\tfrac12}\B{h}_\text{s} \times \im \big[ \B{\alpha} + \underline{\B{\alpha}} \, \big].
\end{equation}
By multiplying \cref{eq:anticomm} by appropriate Pauli matrices, taking traces, and using $\tr[\hat{A}^\dagger] = \tr[\hat{A}]^*$, one can show that ${\underline{\B{\alpha}} = -\B{\alpha}^*}$.
This makes $\B{\alpha} - \underline{\B{\alpha}}$ real and $\B{\alpha} + \underline{\B{\alpha}}$ imaginary, so both contributions are compatible with the normalization condition.
We could also use this information to eliminate the underlined coefficients, but this would make it harder to see how mixed singlet/triplet-terms cancel {later in the derivation}.
Interestingly, all spin supercurrent contributions depend on the same coefficient $\B{\alpha}$, and do not couple to the other traces of $\gr\nabla\gr$.

The physically observable spin supercurrent is found by integrating the spectral current over all positive and negative energies.
We also know that $h_0$ and $\B{h}_\text{s}$ are odd and even functions of energy, respectively.
We can therefore let $\underline{\B{\alpha}}\,(+\epsilon) \rightarrow \mp\underline{\B{\alpha}}\,(-\epsilon) = \mp\underline{\tilde{\B{\alpha}}}{\,}^*(+\epsilon)$ in the spectral current {without changing the total spin supercurrent:}
\begin{equation}
  \label{eq:spincurrent-useful}
  \B{j}_\text{s} = {\tfrac12} h_0     \re \big[ \B{\alpha} + \underline{\tilde{\B{\alpha}}}{\,}^* \, \big]
  + {\tfrac12} \B{h}_\text{s} \times \im \big[ \B{\alpha} + \underline{\tilde{\B{\alpha}}}{\,}^* \, \big].
\end{equation}
This form of the result will be useful later, as it makes it clearer which parts of the non-underlined and underlined coefficients cancel for symmetry reasons.
Conveniently, this also makes the $h_0$ and $\B{h}_\text{s}$ contributions take very similar forms.

\subsection{Expansion in singlets and triplets}
We now proceed with an expansion of the propagators in terms of physically meaningful components.
Following the same kind of parametrization as Ref.~\cite{Eschrig2015a}, we can write
\begin{equation}
  \gr = 
  \begin{pmatrix}
    (g_s + \B{g}_t\cdot\B{\sigma}) & (f_s + \B{f}_t\cdot\B{\sigma}) i\sigma_2 \\
    -i\sigma_2 (\tilde{f}_s - \tilde{\B{f}}_t\cdot\B{\sigma}) & -\sigma_2 (\tilde{g}_s - \tilde{\B{g}}_t\cdot\B{\sigma}) \sigma_2 \\
  \end{pmatrix}.
  \label{eq:singlet-triplet-param}
\end{equation}
{Here, $f_s$ represents the spin-singlet pair amplitude, while $\B{f}_t$ is the spin-triplet amplitude.}
{On the other hand,} we can interpret $g_s$ and $\B{g}_t$ as the spin-independent and spin-dependent parts of the density of states, respectively \cite{Eschrig2015a}.
In our notation, this means that the density of states for particles with spin-projection~$\B{p}$ is given by $N = N_0 \re[g_s + \B{g}_t\cdot\B{p}]$.
In equilibrium, the spin accumulation is found by integrating~$h_0\B{g}_t$ over energies, giving another interpretation of~$\B{g}_t$.
Outside of equilibrium, we of course get another kind of spin accumulation due to a nonzero spin mode~$\B{h}_\text{s}$, which we are interested in here.

Using \cref{eq:singlet-triplet-param} and the identity $\sigma_2\B{\sigma}\sigma_2 = -\B{\sigma}^*$, we find that the diagonal components of $\gr\nabla\gr$ in Nambu space are
\begin{equation}
\begin{aligned}
  \big[ \gr\nabla\gr \big]_{1,1} 
    &= (g_s + \B{g}_t\cdot\B{\sigma}) \nabla (g_s + \B{g}_t\cdot\B{\sigma}) \\
    &\,+ (f_s + \B{f}_t\cdot\B{\sigma}) \nabla (\tilde{f}_s - \tilde{\B{f}}_t\cdot\B{\sigma}), \\
  \big[ \gr\nabla\gr \big]_{2,2} 
    &= (\tilde{g}_s + \tilde{\B{g}}_t\cdot\B{\sigma}^*) \nabla (\tilde{g}_s + \tilde{\B{g}}_t\cdot\B{\sigma}^*) \\
    &\,+ (\tilde{f}_s + \tilde{\B{f}}_t\cdot\B{\sigma}^*) \nabla (f_s - \B{f}_t\cdot\B{\sigma}^*),
\end{aligned}
\end{equation}
{where the subscripts $[\cdots]_{i,j}$ are matrix indices in Nambu space.}
Using the identity $(\B{a}\cdot\B{\sigma})(\B{b}\cdot\B{\sigma}) = (\B{a}\cdot\B{b})+i(\B{a}\times\B{b})\cdot\B{\sigma}$ and its conjugate $(\B{a}\cdot\B{\sigma}^*)(\B{b}\cdot\B{\sigma}^*) = (\B{a}\cdot\B{b})-i(\B{a}\times\B{b})\cdot\B{\sigma}^*$, we can sort the above into spin-independent and spin-dependent terms,
\begin{equation}
\begin{aligned}
  \big[ \gr\nabla\gr \big]_{1,1} 
  &=   \big( g_s \nabla g_s + \B{g}_t \nabla \B{g}_t + f_s \nabla \tilde{f}_s - \B{f}_t \nabla \tilde{\B{f}}_t \big) \\
  &\,+ \big( g_s \nabla \B{g}_t + \B{g}_t \nabla g_s + \B{f}_t \nabla \tilde{f}_s - f_s \nabla \tilde{\B{f}}_t \big)\B{\sigma} \\
  &\,+ \big( \B{g}_t\times\nabla\B{g}_t - \B{f}_t\times\nabla\tilde{\B{f}}_t \big) i\B{\sigma}, \\
  \big[ \gr\nabla\gr \big]_{2,2} 
  &=   \big( \tilde{g}_s \nabla \tilde{g}_s + \tilde{\B{g}}_t \nabla \tilde{\B{g}}_t + \tilde{f}_s \nabla f_s - \tilde{\B{f}}_t \nabla \B{f}_t \big) \\
  &\,+ \big( \tilde{g}_s \nabla \tilde{\B{g}}_t + \tilde{\B{g}}_t \nabla \tilde{g}_s + \tilde{\B{f}}_t \nabla f_s - \tilde{f}_s \nabla \B{f}_t \big)\B{\sigma}^* \\
  &\,- \big( \tilde{\B{g}}_t\times\nabla\tilde{\B{g}}_t - \tilde{\B{f}}_t\times\nabla\B{f}_t \big) i\B{\sigma}^*. \\
\end{aligned}
\end{equation}
Since we define $\spinv = \text{diag}(\B{\sigma},\B{\sigma}^*)$, \cref{eq:trace-coefficients} tells us that the coefficient~$\B{\alpha}$ that we require can be expressed as 
\begin{equation}
  \B{\alpha} = \frac14 \tr\left\{\B{\sigma}[ \gr\nabla\gr \big]_{1,1} -  \B{\sigma}^*[ \gr\nabla\gr \big]_{2,2}\right\}. 
\end{equation}
Together with the expansion of $\gr\nabla\gr$ above, and standard trace identities for Pauli matrices, we then obtain
\begin{equation}
\label{eq:alpha-one}
\begin{aligned}
  2\B{\alpha} &=   g_s \nabla \B{g}_t + \B{g}_t \nabla g_s - \tilde{g}_s \nabla \tilde{\B{g}}_t - \tilde{\B{g}}_t \nabla \tilde{g}_s \\
         &\,+ \B{f}_t \nabla \tilde{f}_s - f_s \nabla \tilde{\B{f}}_t - \tilde{\B{f}}_t \nabla f_s + \tilde{f}_s \nabla \B{f}_t \\
         &\,+ i\B{g}_t\times\nabla\B{g}_t + i \tilde{\B{g}}_t\times\nabla\tilde{\B{g}}_t \\
         &\,- i\B{f}_t\times\nabla\tilde{\B{f}}_t - i\tilde{\B{f}}_t\times\nabla\B{f}_t .\\
\end{aligned}
\end{equation}

Let us now calculate the corresponding coefficient $\underline{\B{\alpha}}$ from the matrix $\gr^\dagger\nabla\gr^\dagger$.
Taking the complex-transpose of \cref{eq:singlet-triplet-param}, 
\begin{equation}
  \gr^\dagger = 
  \begin{pmatrix}
    (g_s^* + \B{g}_t^*\cdot\B{\sigma}) & (\tilde{f}_s^* - \tilde{\B{f}}_t^*\cdot\B{\sigma}) i\sigma_2 \\
    -i\sigma_2 (f_s^* + \B{f}_t^*\cdot\B{\sigma})  & -\sigma_2 (\tilde{g}_s^* - \tilde{\B{g}}_t^*\cdot\B{\sigma}) \sigma_2 \\
  \end{pmatrix},
\end{equation}\\
we see that $\gr^\dagger$ changed as follows compared to $\gr$:
\begin{equation}
\begin{aligned}
  g_s     &\rightarrow +g_s^*, &
  \B{g}_t &\rightarrow +\B{g}_t^*, &
  f_s     &\rightarrow +\tilde{f}_s^*, &
  \B{f}_t &\rightarrow -\tilde{\B{f}}_t^*. 
\end{aligned}
\end{equation}
Other than these transformations, the parametrization is clearly identical, and the derivation of $\underline{\B{\alpha}}$ becomes identical as well.
If we in the end results also choose to let $\epsilon \rightarrow -\epsilon$, corresponding to a combination of complex-conjugation and tilde-conjugation, {the net} transformation rules become
\begin{equation}
\begin{aligned}
  g_s     &\rightarrow +\tilde{g}_s, &
  \B{g}_t &\rightarrow +\tilde{\B{g}}_t, &
  f_s     &\rightarrow +f_s, &
  \B{f}_t &\rightarrow -\B{f}_t. 
\end{aligned}
\end{equation}
We can therefore simply perform the changes above to \cref{eq:alpha-one} to get the corresponding equations for $\underline{\tilde{\B{\alpha}}}{\,}^*$:
\begin{equation}
\label{eq:alpha-two}
\begin{aligned}
  2\underline{\tilde{\B{\alpha}}}{\,}^* 
  &=   \tilde{g}_s \nabla \tilde{\B{g}}_t + \tilde{\B{g}}_t \nabla \tilde{g}_s - g_s \nabla \B{g}_t - \B{g}_t \nabla g_s \\
  &\,- \B{f}_t \nabla \tilde{f}_s + f_s \nabla \tilde{\B{f}}_t + \tilde{\B{f}}_t \nabla f_s - \tilde{f}_s \nabla \B{f}_t \\
  &\,+ i\tilde{\B{g}}_t\times\nabla\tilde{\B{g}}_t + i \B{g}_t\times\nabla\B{g}_t  \\
  &\,- i\B{f}_t\times\nabla\tilde{\B{f}}_t - i\tilde{\B{f}}_t\times\nabla\B{f}_t.\\
\end{aligned}
\end{equation}

We are now ready to calculate the spectral spin supercurrent in terms of the singlet/triplet-decomposition.
Adding up \cref{eq:alpha-one,eq:alpha-two}, we see that all mixed {singlet/triplet} terms drop out, and we are left with only the cross-product terms:
\begin{equation}
  \label{eq:alpha-final}
  \B{\alpha} + \underline{\tilde{\B{\alpha}}}{\,}^* =  
  + i \B{g}_t\times\nabla\B{g}_t + i\tilde{\B{g}}_t\times\nabla\tilde{\B{g}}_t 
  - i\tilde{\B{f}}_t\times\nabla\B{f}_t - i\B{f}_t\times\nabla\tilde{\B{f}}_t .
\end{equation}
Substituting this into \cref{eq:spincurrent-useful}, we immediately see that:
\begin{equation}
  \begin{aligned}
    \B{j}_\text{s} 
    = - {\tfrac12} h_0            &\im\big[\B{g}_t\times\nabla\B{g}_t + \tilde{\B{g}}_t\times\nabla\tilde{\B{g}}_t -\B{f}_t\times\nabla\tilde{\B{f}}_t -\tilde{\B{f}}_t \times \nabla\B{f}_t \big] \\
    +\, {\tfrac12} \B{h}_\text{s} \times &\re\big[ \B{g}_t\times\nabla\B{g}_t + \tilde{\B{g}}_t\times\nabla\tilde{\B{g}}_t -\B{f}_t\times\nabla\tilde{\B{f}}_t -\tilde{\B{f}}_t \times \nabla\B{f}_t \big].
  \end{aligned}
\end{equation}
Since $h_0$ and $\B{h}_\text{s}$ are odd and even functions of energy, and the observable spin current is the energy integral of the above, we can let $h_0 \im[\tilde{A}] \rightarrow h_0 \im[A]$ and $\B{h}_\text{s} \re[\tilde{A}] \rightarrow \B{h}_\text{s} \re[A]$ without changing any results.
Applied to the above, we can summarize our results in the tidy and compact form
\begin{equation}
  \label{eq:final-result}
  \begin{aligned}
    \B{j}_\text{s} 
    = &\, {-\im}\big[\B{g}_t\times\nabla\B{g}_t - \B{f}_t\times\nabla\tilde{\B{f}}_t \big] \kern0.06em\cdot\kern0.06em h_0 \\
      &\, {-\re}\big[\B{g}_t\times\nabla\B{g}_t - \B{f}_t\times\nabla\tilde{\B{f}}_t \big] \times \B{h}_\text{s}.
  \end{aligned}
\end{equation}
We have shown earlier in the derivation that both contributions are compatible with the normalization condition.
The fact that they {did not} cancel during the last simplification above, shows that both contributions are compatible with the energy symmetries of $h_0$ and $\B{h}_\text{s}$.
Finally, we know that {the contents of the brackets} $\B{g}_t\times\nabla\B{g}_t - \B{f}_t\times\nabla\tilde{\B{f}}_t$ can be nonzero, since this is the source of equilibrium spin currents.

{The final result shows that if one in equilibrium has a spin supercurrent~$\B{j}_{\text{s}}^\text{eq}$, then a nonequilibrium spin mode~$\B{h}_\text{s}$ gives rise to a new component~$\B{j}_\text{s}^\text{neq} \sim \B{j}_{\text{s}}^\text{eq}\times\B{h}_\text{s}$.
This can intuitively be interpreted as the injected spins $\B{h}_\text{s}$ exerting some kind of torque on the spins transported by the equilibrium current~$\B{j}_\text{s}^\text{eq}$, thus producing a component $\B{j}_{\text{s}}^\text{neq}$ that is spin-polarized in a direction perpendicular to both.
This analogy is not perfect: it leaves out the Im and Re operations in \cref{eq:final-result}, and the fact that the cross-product relation is between \emph{spectral} currents and accumulations.
However, the intuition provided by this picture is sufficient to explain the results in the main manuscript.}

\clearpage
\section{Numerical model}\label{sec:numerics}
As summarized in the main article, we perform the numerical calculations using the Usadel formalism~\cite{BergeretArxiv,Chandrasekhar2008,Belzig1999,Rammer1986,Usadel1970}.
This is formulated in terms of the $8\times8$ quasiclassical propagator
\begin{equation}
  \gv \coloneq
  \begin{pmatrix}
    \gr & \gk \\
    0   & \ga
  \end{pmatrix},
\end{equation}
which satisfies the identities $\gk = \gr \gh - \gh \ga$ and $\ga = \H{\tau}_3 \gr^\dagger \H{\tau}_3$.
Together, these identities show that we have to determine two $4\times4$ matrices to know~$\gv$: the retarded propagator~$\gr$, which determines the spectral properties of a material; and the distribution function~$\gh$, which describes the occupation numbers of the states in the material.
Both of these matrices can be functions of position~$\B{r}$ and quasiparticle energy~$\epsilon$.

In general, the distribution function~$\gh$ follows from solving a \emph{kinetic equation} that can be derived directly from the full $8\times8$ Usadel equation.
We present a complete derivation of a kinetic equation and relevant boundary conditions in Ref.~\cite{OuassouArxivA}, which is valid for quite general superconducting structures.
The result is formulated as an explicit and linear differential equation, which can be easily and efficiently implemented in a numerical Usadel solver.
Related derivations can be found in Refs.~\cite{Bobkova2015a, Silaev2015, BergeretArxiv, Aikebaier2018, Chandrasekhar2008, Belzig1999}.
However, instead of solving the kinetic equation explicitly, we have made two simplifying assumptions about the distribution function~$\gh$.
The first is that it is roughly constant throughout the superconductor, which is reasonable as long as the superconductor is not too thick compared to its spin relaxation length.
The second is that the distribution function can be modelled using a spin voltage, which we justify in the Discussion in the main article.
These assumptions imply that we can treat $\gh$ as a constant parameter, which simplifies our model system from a 2D to 1D geometry, thus making it much more feasible to attack numerically.

Since we treat $\gh$ as a parameter, we only have to solve the Usadel equation for $\gr$.
In an effectively 1D superconductor, such as the model system in the main article, this takes the form
\begin{equation}
  \label{eq:supp-usadel-eq}
  i\xi^2\partial_x(\gr\partial_x\gr) = [\H{\Delta} + \epsilon\H{\tau}_3 ,\, \gr]/\Delta_0,
\end{equation}
where $\H{\Delta} = \text{antidiag}(+\Delta,-\Delta,+\cc{\Delta},-\cc{\Delta})$ incorporates the superconducting order parameter $\Delta(x)$, $\H{\tau}_3 = \text{diag}(+1,+1,-1,-1)$ is a Pauli matrix in Nambu space, $\Delta_0$ is the bulk zero-temperature order parameter, and $\xi$ is the superconducting coherence length.
In the numerical simulations, we also approximate the effect of inelastic scattering using a Dynes parameter: $\epsilon \mapsto \epsilon+0.01i\Delta_0$.

The ferromagnetic insulators in our model system were treated as spin-active interfaces, \ie boundary conditions that account for spin-dependent phase shifts for quasiparticles in 
\newpage 
\noindent
the superconductor that are reflected from the insulating interface~\cite{Eschrig2015b, Machon2013, Cottet2009, Cottet2007a}.
This boundary condition takes the form
\begin{equation}
  \label{eq:supp-usadel-bc}
  \pm 2L (\gr \partial_x \gr) = -i(G_\varphi/G_\textsc{n})\, [\B{m}\cdot\H{\B{\sigma}}, \gr],
\end{equation}
where the plus and minus signs describe boundary conditions at the left and right interfaces of the superconductor, respectively.
Here, $L$ is the length of the superconductor along the $x$-axis, $G_\textsc{n}$ is the bulk conductance of the superconductor in its non-superconducting state, $G_\varphi$ describes the effect of spin-dependent phase shifts when quasiparticles are reflected from a magnetic interface, and $\B{m}$ is a unit vector that describes the magnetization direction at the same interface.

In addition to the equations above, we need to solve a self-consistency equation for the order parameter~$\Delta(x)$~\cite{Jacobsen2015a},
\begin{equation}
  \label{eq:supp-gap-sc}
  \Delta(x) = \frac{1}{4 \log(2\omega_\text{c}/\Delta_0)} \int_{-\omega_\text{c}}^{+\omega_\text{c}} \d{\epsilon}\, f^{\textrm{\smaller K}}_{s}(x, \epsilon).
\end{equation}
Here, $f^{\textrm{\smaller K}}_{s}$ is the singlet part of the anomalous part of the Keldysh propagator~$\gk$.
In practice, it can be evaluated at all energies from the calculated $\gk$ at positive energies using the identities
\begin{align}
  f^{\textrm{\smaller K}}_{s}(x,+\epsilon) &= \frac{1}{4}\tr\big[ (-i\H{\sigma}_2)\, (\H{\tau}_1 - i\H{\tau}_2)\, \gk(x,+\epsilon)\big],        \label{eq:supp-gap-id1}\\
  f^{\textrm{\smaller K}}_{s}(x,-\epsilon) &= \frac{1}{4}\tr\big[ (+i\H{\sigma}_2)\, (\H{\tau}_1 + i\H{\tau}_2)\, \gk(x,+\epsilon)\big]^*\!\!.   \label{eq:supp-gap-id2}
\end{align}
These follow from parametrizing $\gk$ in a similar manner as \cref{eq:singlet-triplet-param}, then invoking the definition of tilde-conjugation $\tilde{f}_s(x,\epsilon) \coloneq f_s^*(x,-\epsilon)$, and finally using standard trace identities.

The actual numerical implementation was done using the Riccati parametrization of the retarded propagator~\cite{SchopohlArxiv,Jacobsen2015a},
\begin{align}
  \gr = 
  \begin{pmatrix}
    +N & 0 \\
    0 & -\tilde{N}
  \end{pmatrix}
  \begin{pmatrix}
    1+\gamma\tilde{\gamma} & 2\gamma \\
    2\tilde{\gamma} & 1+\tilde{\gamma}\gamma
  \end{pmatrix},
\end{align}
where $\gamma$ is a $2\times2$ complex matrix, and $N \coloneq (1-\gamma\tilde{\gamma})^{-1}$.
This matrix structure is defined in a way that automatically satisfies the normalization condition~$(\gr)^2 = 1$, and accounts for the particle--hole symmetries of the propagator, thus reducing the number of independent variables one has to solve for numerically.
This parametrization also has the additional benefits that the Riccati parameter~$\gamma$ is single-valued and has a bounded norm, thus resulting in a more stable numerical solution procedure.
How to reformulate the equations for $\gr$ in terms of $\gamma$ is described in \eg Ref.~\cite{Jacobsen2015a}.
In practice, we alternate between calculating $\gamma$ from \cref{eq:supp-usadel-eq,eq:supp-usadel-bc}, and updating $\Delta(x)$ using \cref{eq:supp-gap-sc,eq:supp-gap-id1,eq:supp-gap-id2}, until they converge to a satisfactory degree.
The simulation code itself is publicly available from \href{https://github.com/jabirali/GENEUS}{github.com/jabirali/\textsc{geneus}}.

\clearpage

%-------------------------------------------------------------------------------%
%                                  BIBLIOGRAPHY                                 %
%-------------------------------------------------------------------------------%

%\bibliography{references}

% Bibtex preamble
%